\documentclass[12pt]{iopart}

\usepackage[latin1]{inputenc}
\usepackage{graphicx}
\usepackage{color,textcomp}
\begin{document}
\title{Time reversal of light by linear dispersive filtering near atomic resonance}
\author{ H Linget, T Chaneli\`ere, J-L Le Gou\"et and  A Louchet-Chauvet}
\address{Laboratoire Aim\'{e} Cotton, CNRS UPR3321, Univ. Paris Sud, b\^atiment 505,
campus universitaire, 91405 Orsay, France}
\ead{anne.chauvet@u-psud.fr}
\begin{abstract}
Based on the similarity of paraxial diffraction and dispersion mathematical descriptions, the temporal imaging of optical pulses combines linear dispersive filters and quadratic phase modulations operating as time lenses. We consider programming a dispersive filter near atomic resonance in rare earth ion doped crystals, which leads to unprecedented high values of dispersive power. This filter is used in an approximate imaging scheme, combining a single time lens and a single dispersive section and operating as a time reversing device, with potential applications in radio-frequency signal processing. This scheme is closely related to three-pulse photon echo with chirped pulses but the connection with temporal imaging and dispersive filtering emphasizes new features.
\end{abstract}
\date{\today}

\section{Introduction}
The time reversal of a light pulse obeys a very general condition imposed by causality. Namely, in order to be ultimately time-reversed, the entire light pulse has to be shelved within the time reversing device at some moment. Indeed, no radiative emission should be expected until the pulse tail, to be converted into the time-reversed pulse head, has been captured. One cannot help noticing some similarities with other actively investigated problems, such as slow light and optical memories.

In addition to the scientific implications suggested by this general context, the time reversal of light also offers stimulating technological prospects. As initially demonstrated in the acoustic wave domain, space phase conjugation has to be combined with time reversal to concentrate pulsed energy. This has been applied to medical lithotripsy techniques~\cite{fink1997}. In the framework of electronic warfare one is faced with similar challenges in the radio-frequency range. If time reversal is conveniently achieved at low frequencies by electronic sampling, recording and playing back~\cite{lerosey2004}, the broadband regime requires alternative approaches, such as all-optical processing of an upconverted radio-frequency waveform.

In the present paper we analyze time reversal in the framework of Fourier optics, taking advantage of the similarity of linear group delay dispersion and paraxial diffraction. Within this picture, we disclose an approximate temporal imaging scheme that time reverses the incoming waveform in the same way as a lens generates an inverted image. This scheme is considered for time-reversal of optically-carried radio-frequency signals.

As pointed out above, the waveform to be time-reversed has to be entirely shelved in the device at some moment. The condition is handled quite easily with pulses shorter than a few nanoseconds. The task is more challenging when the waveform duration reaches a few microseconds. To manage this issue, we use a linear dispersive filter working in the vicinity of atomic resonance. The waveform is saved in long-lived atomic superposition states, just enough time until the signal tail is captured and launched back as the head of the reversed light pulse.

In Section~\ref{sec:general_framework}, we place time reversal within the general framework of temporal imaging. After a rapid historical survey we consider an approximate imaging scheme that gives rather simple access to time reversal. We point out the connection with the ancient pinhole camera. In Section~\ref{sec:shaped absorption} we consider the devising of the linear dispersive filters required for temporal imaging. We show that such filtering can be provided by appropriately programmed media near atomic resonances. Experimental results are presented in Section~\ref{sec:experimental}.

\section{Temporal imaging and time reversal}\label{sec:general_framework}
\subsection{Fourier optics and temporal imaging}\label{sec:temporal_imaging}
With the advent of microwave and optical coherent sources, Fourier optics was actively developed in the sixties~\cite{goodman}. Quite soon it was noticed that the paraxial diffraction of a spatial wavefront obeys the same mathematical description as the propagation of a time domain waveform through a dispersive medium. Both descriptions rely on Fresnel transforms that just differ by the transform variable. The transverse space coordinate of diffraction is replaced by time in dispersion. This time-space duality was first pointed out in the framework of RADAR pulse compression~\cite{tournois64,tournois68}. The extension to optical domain~\cite{treacy69} was related to the rapid progress of laser pulse compression in the femtosecond range~\cite{grischkowsky74,moll1980,shank1982,tomlinson1984}.

Typically, pulse compression involves two steps. First, additional spectral components are generated by a non-linear process such as self-phase modulation. Ideally this is equivalent to "chirping" the pulse frequency, $i.e.$ modulating the incoming field with the quadratic phase factor $\mathrm{e}^{irt^2/2}$, where $r$ stands for the chirp rate. While preserving the temporal profile, the transformation broadens the spectrum, displaying the different spectral components in a time-ordered sequence. In a second step, the time separated spectral terms are synchronized by linear filtering through a matched dispersive element, which results in pulse compression. The dispersive filter multiplies the spectral component at $\omega$ by the quadratic phase factor $\exp i\phi(\omega)$ where $\phi(\omega)=\mu\omega^2/2$. The dispersion stage is matched to the chirping step if $\mu r=1$. Then the emerging waveform reads as $\tilde{\cal{E}}(rt)\mathrm{e}^{irt^2/2}$, where $\tilde{\cal{E}}(\omega)$ represents the time-to-frequency Fourier transform of the incoming pulse $\cal{E}$$(t)$. Hence the waveform spectrum is displayed in the time domain at the filter output. Compression can occur provided $1/r$ is smaller than the squared duration of $\cal{E}$$(t)$.

The frequency chirping step generates spectral components in the same way as a conventional lens adds wavevectors to the spatial wavefront. The time-domain phase shift $rt^2/2$ is reminiscent to $\pi(x^2+y^2)/\lambda f$, the phase shift that affects a wavefront at wavelength $\lambda$, along the path at transverse position $(x,y)$ through a conventional thin lens with focal distance $f$. Then, the dispersive filtering step with $\mu r=1$ is equivalent to the free space diffraction of a wavefront from the lens to the focal plane. The two-step process performs the Fourier transform of the incoming waveform in the same way as a lens takes the far field or Fraunhofer diffraction pattern of an incident wavefront into the focal plane~\cite{Jannson-OL83}.

Therefore, time space duality in optics is not limited to the similarity of diffraction and dispersion. It also extends to the lensing function. Actually the mathematical analysis of spatial imaging can be applied to time-domain waveforms, and any coherent optics element can be assigned a counterpart in the time domain. From this emerges the concept of temporal imaging~\cite{kolner89,kolner94}.

Temporal imaging involves three steps at least. The incoming waveform first passes through a dispersive element, with dispersion coefficient $\mu_1$, that operates as a region of free-space Fresnel diffraction in the space domain. Then the waveform is frequency chirped by factor $\mathrm{e}^{irt^2/2}$, which is similar to the action of a thin lens on the transverse spatial profile of a beam. Finally, the waveform passes through a second dispersive region, with dispersion coefficient $\mu_2$, equivalent to the lens-to-image free-space propagation in space-domain optics. Provided dispersion in the input and output sections is correctly matched to the degree of phase modulation by the time lens, a time-reversed replica of the input waveform is received at the output. The matching condition, $1/\mu_1+1/\mu_2=r$, is similar to the thin lens conjugation formula. The stretching factor $-\mu_2/\mu_1$, depending on the relative strength of the two dispersive sections, replaces the space domain magnification factor~\cite{bennett94}. Just as in the space domain, a single lens correctly images the object intensity distribution, but alters the phase. A two-lens telescope~\cite{foster09} is needed to properly image the field distribution.

Although most effort has been concentrated for a long time on exactly matched imaging or Fourier transform schemes, a new interest has arisen quite recently in situations where the Fraunhofer condition is only approximately satisfied. The approximate Fourier transform capabilities of a simple dispersion line was first identified in the context of real-time spectral analysis~\cite{muriel1999,azana2000}. It was pointed out that, although no lens was used to bring the far field zone to finite distance, the Fraunhofer condition could be satisfied provided the squared duration of the incoming pulse is smaller than the line dispersion coefficient $\mu$. This procedure, known as wavelength-to-time mapping, was then extended to arbitrary waveform generation~\cite{chou2003,lin2005} and to approximate imaging~\cite{coppinger99,azana2005}.

\subsection{Approximate imaging for time reversal}\label{sec:approximate_imaging}
In many temporal imaging demonstrations, the lens effect is obtained by mixing the beam carrying the waveform with a frequency chirped pulse in a two-wave or four-wave mixing non-linear material~\cite{bennett94,bennett99,bennett00,foster09}. Without resorting to parametric optical processes, one can implement the lens effect in a simpler way by just modulating the amplitude of a frequency chirped optical carrier. The price to pay is that the optically carried waveform is created by the lens itself. In other words, the optical signal does not exist before shaping by the lens. This rules out the input step of propagation through a dispersive region, required in the temporal imaging scheme (see section~\ref{sec:temporal_imaging}). Modulating the amplitude of a frequency chirped carrier is equivalent to having a collimated beam pass through an object attached to a lens. Combining this input step with a single dispersive section, one can build an approximate time-reversed image of the incoming waveform.

\begin{figure}[!htb]
\centering\includegraphics[width=8.5cm]{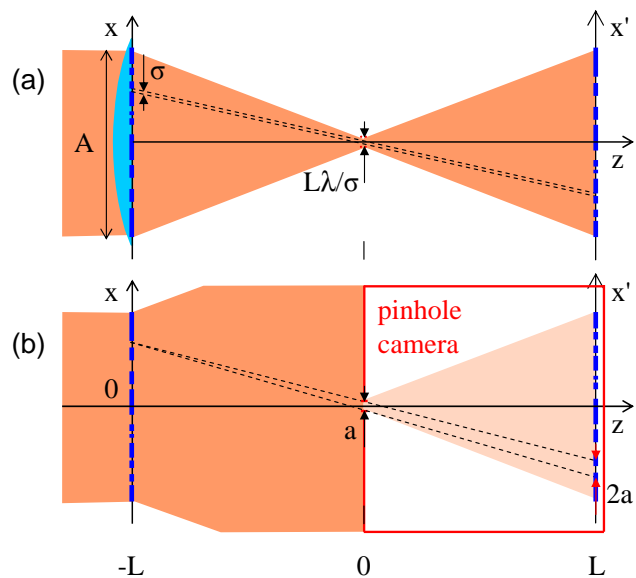}\caption{Conventional optics image reversal schemes. In upper box (a), the object at $z=-L$ is illuminated by a collimated beam. The size of the object smallest detail is represented by $\sigma$. The lens, placed side-by-side with the object, concentrates the energy on a $\lambda L/\sigma$-wide bottleneck in the focal plane ($z=0$). The imaging spatial resolution on the screen at $z=L$, of order $\sqrt{\lambda L}$, is limited by the Fraunhofer condition. The lower box (b) represents the pinhole camera. The black dashed lines depict the camera impulse response. They represent geometrical rays that are traced from a single point on the object to the screen, through the $a$-wide slit. The shaded zones with different shade-levels are intended to show that a small part of the incoming energy is allowed by the slit to the screen.}
\label{fig:space_pinhole}%
\end{figure}

Before considering operation in the time domain, let us first address the conventional optics counterpart, as sketched in figure~\ref{fig:space_pinhole}(a). An $A$-wide object $E(x)$ is positioned at $z=-L$ against a converging lens with focal distance $L$. Let $\sigma\ll A$ represent the smallest characteristic distance of variation of the object transmission. After crossing the lens and the object, an incoming collimated beam travels to a screen located at $z=L$. A detailed description of the propagation of a spherical wave after passing through an object can be found in~\cite{azana2005shadow}. We choose a different but complementary point of view and analyze the field propagation from $z=-L$ to $z=L$ in two segments.

In the first segment, from $z=-L$ to $z=0$, the wavefront emerging from the lens is converted into the far field or Fraunhofer diffraction pattern, brought to finite distance, at $z=0$, by the lens. The focal spot, of size $\lambda L/\sigma$, is nothing but the Fourier transform $\widetilde{E}(u/(\lambda L))$ of the incoming signal $E(x)$, multiplied by the phase factor $\mathrm{e}^{i\pi u^2/(\lambda L)}$.

Traveling beyond the focal plane, the wavefront enters the second segment where free space propagation again operates as a Fourier transformer in the Fraunhofer limit. From $z=0$ to $z=L$, the Fraunhofer condition is satisfied provided the squared spot size $(\lambda L/\sigma)^2$ at $z=0$ is smaller than $\lambda L$, which leads to $\lambda L<\sigma^2$. After undergoing two successive Fourier transforms, the intensity transverse distribution at $z=L$, equals $|E(-x')|^2$. According to the Fraunhofer condition and to the $\sigma\ll A$ assumption, the $\lambda L/\sigma$-wide focal spot behaves as a narrow bottleneck, much smaller than the object size, through which all the incoming energy is channelized. The $\lambda L<\sigma^2$ condition sets the distortion-less imaging capabilities of the device.

Back to the time domain, let the incoming signal $S(t)$ be injected in the dispersive medium after being mixed with the optical carrier by the time lens. When the dispersive line coefficient, $\mu$, equals the lens inverse chirp rate $1/r$, the signal in the "focal plane" of the time lens is nothing but the time-to-frequency Fourier transform $\widetilde{S}(\omega)$ of $S(t)$, in accordance with the above conventional optics discussion. Specifically, the "focal plane" signal $S_\mathrm{foc}(t)$ reads as $\tilde{S}(rt)\mathrm{e}^{irt^2/2}$.

\begin{figure}[!htb]
\centering\includegraphics[width=7.5cm]{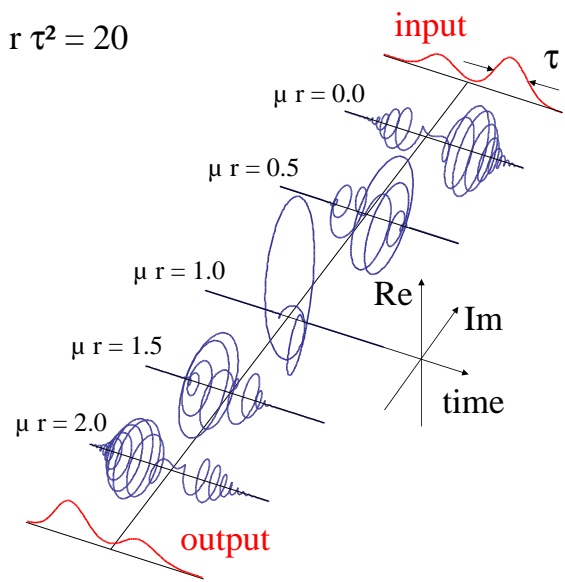}\caption{Approximate imaging for time reversal. Passing through a time lens, the incoming $S(t)$ waveform is changed into $S_{\mu=0}(t)=S(t)\mathrm{e}^{irt^2/2}$. Then $S_\mu(t)$ propagates through a dispersive filter, from $\mu=0$ at the input to $\mu=2/r$ at the output. The time variation of $S_\mu(t)$ is displayed in 3D representation at different $\mu$-positions, illustrating the waveform compression into the focal plane, at $\mu=1/r$, the subsequent stretching and the time-reversal at $\mu=2/r$. The phase evolution in the focal plane results from the time separation of the two incoming peaks. The output profile represents $\left|S_{\mu=2/r}(t)\right|$. In this simulation, $r$ has been set equal to $20/\tau^2$, where $\tau$ represents the full width at half maximum of the incoming peaks. }
\label{fig:camera_obscura}%
\end{figure}

If $\mu>1/r$, the waveform travels beyond the "focal plane". Just as free space propagation in conventional optics, the dispersive line section downstream from the focal plane, operates again as a Fourier transformer in the Fraunhofer limit~\cite{muriel1999,azana2000,azana2005}. When $\mu=2/r$, the dispersion coefficient of this section reads $1/r$, and the Fraunhofer condition is satisfied if the squared duration of $S_\mathrm{foc}(t)=\tilde{S}(rt)\mathrm{e}^{irt^2/2}$ is smaller than $1/r$. The duration of $S_\mathrm{foc}(t)$ equals $(r\sigma)^{-1}$, where $\sigma$ represents the smallest characteristic variation time interval of $S(t)$. Hence the Fraunhofer condition reads as $r\sigma^2>1$ and $\mathrm{e}^{irt^2/2}\approx 1$ over the width of $S_\mathrm{foc}(t)$. Finally, after undergoing a pair of Fourier transforms, the time-reversed waveform $S(-t)$ is displayed at the output~\cite{coppinger99} (see figure~\ref{fig:camera_obscura}).

The image sharpness condition can also be given a group delay picture. Let $\omega_0$ be an arbitrary frequency within the dispersive filter frequency bandwidth. At a given frequency $\omega$, input and output are separated by the group delay $\tau_g(\omega)=\mathrm{d}_\omega\phi(\omega)$.  At the line output, where $\mu=2/r$,  $\tau_g(\omega)=\tau_g(\omega_0)-2(\omega-\omega_0)/r$. Let a Fourier-transform-limited, $\tau$-long, temporal substructure be injected at frequency $\omega$ in the dispersive medium. This pulse spreads over a spectral interval of order $1/\tau$. Due to group delay dispersion $\mathrm{d}_\omega\tau_g(\omega)=-2/r$, the pulse undergoes stretching of order $|\mathrm{d}_\omega\tau_g(\omega)|/\tau=2/(r\tau)$ while travelling through the dispersive filter, and preserves its initial duration provided $\tau>2/(r\tau)$. Hence, $\sqrt{2/r}$ represents the duration of the shortest substructure that can propagate throughout the medium without distortion. The proper time-reversal bandwidth is thus limited to $\sqrt{r/2}$.

Although attractive in the time domain, this scheme is never considered in conventional optics. Indeed, as soon as a lens is available, one takes care of conjugating the object and the image through the lens, which relaxes the distortion-less bandwidth condition. However, the depicted approximate imaging design is rather reminiscent to the pinhole camera, an ancient lens-less imaging device.

\subsection{Connection with the pinhole camera}
To draw the parallel with the pinhole camera, we return to conventional, space domain, optics. In the above description we pointed out that the focal spot can be regarded as a narrow bottleneck through which all the wavefront information is conveyed.

Channelizing through a narrow opening obviously evokes the pinhole camera - a light-proof box with a small hole on one side. Light from a scene enters through the tight opening and projects an inverted image on the opposite side of the box [see figure~\ref{fig:space_pinhole}(b)]. The pinhole camera operates in Fraunhofer conditions, just as the device shown in figure~\ref{fig:space_pinhole}(a), and offers the same distortion-less resolution. To satisfy the Fraunhofer condition, the opening should be smaller than $\approx\sqrt{\lambda L}$. To minimize the low-pass filtering of angular frequencies, one should not reduce the opening size beneath this value that appears to be the best trade-off between the Fraunhofer and the diffraction limits.

Therefore the two devices offer the same distortion-less resolution, as imposed by the Fraunhofer condition. However, despite their similar geometry, they do not share any additional common property.

First, the Fraunhofer condition is always satisfied inside the pinhole box, as soon as the opening is properly sized. When object details are smaller than $\approx\sqrt{\lambda L}$, they give rise to high angular frequency components that are just cut-off by the hole. In the image on the back side of the box, the details are simply blurred by this low-pass filtering effect. In the figure~\ref{fig:space_pinhole}(a) device, the high angular frequencies are not cut but they violate the Fraunhofer condition. We shall examine this point in depth further on.

Second, the pinhole camera poorly operates in coherent illumination conditions. Indeed, if a $\sigma$-wide detail on the object, located at distance $x$ from the hole axis, is illuminated by a collimated coherent beam, the transmitted light travels in forward direction within a diffraction limited aperture. Hence it reaches the hole only if the $\approx\lambda L/\sigma$-wide projection is smaller than $x$. Otherwise, the rays are just stopped by the light-proof wall of the box. Hence, only a small fraction of the object, very close to the hole-axis, can contribute to the image on the back side of the box. This issue can be solved by spatially incoherent illumination, with a correlation distance along $Ox$ much smaller than $\sigma$. This sub-modulation gives rise to the angular frequencies that are needed to have some light reach the hole from any point of the scene. However, the incoherent illumination of the scene generates angular frequencies in a random way, so that most of the incoming light is lost, as illustrated in figure~\ref{fig:space_pinhole}(b). On the contrary, the figure~\ref{fig:space_pinhole}(a)-device correctly works with coherent light since the lens generates the proper angular frequencies to direct all the light through the bottleneck.

In a time domain pinhole camera, the waveform to be time-reversed would modulate a broadband carrier, with an autocorrelation time much smaller than the waveform shortest temporal detail. A short temporal gate would play the part of the pinhole, through which dispersed components of the initial waveform would be transmitted~\cite{kolner97}.

\section{Time reversal through a shaped-absorption medium}\label{sec:shaped absorption}
\subsection{Linear dispersive filtering near atomic resonance}\label{filter_in_abs_medium}
\label{sec:dispersive_filter}
Any transparent medium can be used as a dispersive filter. Since the dispersive effect grows with the material thickness, hundred-of-meter-long optical fibers can be regarded as emblematic media for such applications. Specific dispersive properties are usually intrinsic to a given material. Here we consider a different class of dispersive filters that are programmed in an absorbing material.

\begin{figure}[tb]
\centering\includegraphics[width=8.5cm]{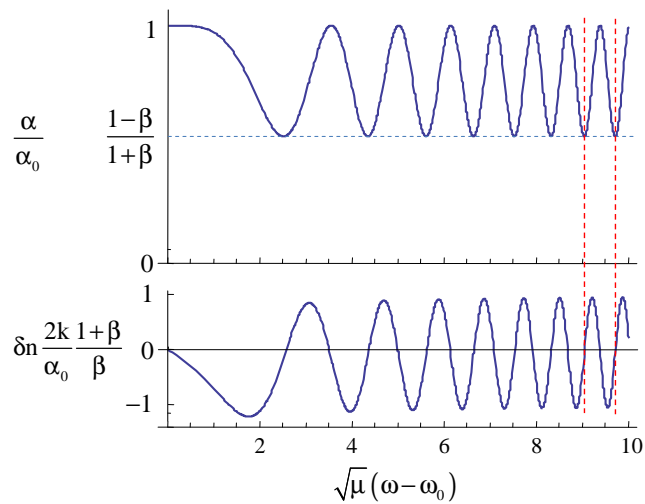}\caption{Spectral profile of the absorption coefficient as defined by equation(~\ref{eq:def_alpha}) (upper box) and of the corresponding index of refraction variation $\delta n(\omega)=n(\omega)-n(\omega_0)$ (lower box). When $\sqrt{\mu}(\omega-\omega_0)$ is large enough, $\alpha(\omega)$ and $\delta n(\omega)$ exhibit the same quasi-sinusoidal shape with the same period and oscillate in quadrature.}
\label{fig:chirped_filter}%
\end{figure}

Let the absorption coefficient of a linear filter be given the following spectral shape:
\begin{equation}
\label{eq:def_alpha}
\alpha(\omega)=\alpha_0\left(1+\beta\cos\left[\phi(\omega-\omega_0)\right]\right)/(1+\beta)
\end{equation}
where $0\leq\beta\leq 1$, $\omega_0$ is an arbitrary frequency origin, and:
\begin{equation}
\label{eq:def_phi}
\phi(\omega)=\mu\omega^2/2
\end{equation}
with $\mu\geq 0$. According to equation~(\ref{eq:def_alpha}), $\alpha(\omega)$ oscillates non-periodically between $\alpha_0$ and $\alpha_0(1-\beta)/(1+\beta)$, as illustrated in figure~\ref{fig:chirped_filter}. The spectral shape is uniformly engraved throughout the depth of the filtering medium. Hence $\alpha(\omega)$ does not depend on the depth coordinate $z$.

Let the field $\mathcal{E}(z=0,\omega)$ be incoming into the filtering material at position $z=0$. At arbitrary depth $z$, the transmission factor reads as:
\begin{equation}
\label{eq:transmission}
\mathcal{T}(z,\omega)=\mathrm{e}^{-ikn(\omega)z-\frac{1}{2}\alpha(\omega)z}=\mathrm{e}^{-ik\left[n(\omega_0)+\frac{1}{2}\chi(\omega)\right]z}
\end{equation}
where $n(\omega)$ and $\chi(\omega)$ respectively represent the index of refraction and the electric susceptibility. Those quantities are linked together by the Kramers-Kr\"onig relations that express causality. Term-by-term identification of the exponents in equation~(\ref{eq:transmission}) leads to:
\begin{equation}
\chi(\omega)=2\left[n(\omega)-n(\omega_0)\right]-\frac{i}{k}\alpha(\omega)
\end{equation}
According to the Kramers-Kr\"onig relations:
\begin{equation}
n(\omega)-n(\omega_0)=-\frac{1}{2k}H\{\alpha\}(\omega)
\end{equation}
where the Hilbert transform $H\{f\}(x)$ is defined as:
\begin{equation}
H\{f\}(x)=\frac{1}{\pi}\mathrm{P}\int_{-\infty}^{\infty}\frac{f(x')}{x-x'}dx'
\end{equation}
One easily shows that, provided $\phi(\omega-\omega_0)\gg1$:
\begin{equation}
H\left\{\mathrm{e}^{\pm i\phi(\omega-\omega_0)}\right\}(\omega)=\mp i\ \mathrm{e}^{\pm i\phi(\omega-\omega_0)},
\end{equation}
which leads to:
\begin{equation}
\label{eq:chi_final}
\chi(\omega)=-\frac{i}{k}\frac{\alpha_0}{1+\beta}\left[1+\beta\mathrm{e}^{-i\phi(\omega-\omega_0)}\right]
\end{equation}
Substituting equation~(\ref{eq:chi_final}) into equation~(\ref{eq:transmission}) and expanding the exponential in series one obtains:
\begin{equation}
\mathcal{T}(z,\omega)=\sum_{m=0}^\infty\mathcal{T}_m(z,\omega)
\end{equation}
where
\begin{equation}
\label{eq:T_n}
\mathcal{T}_m(z,\omega)=\frac{1}{m!}\left(-\frac{1}{2}\frac{\beta\alpha_0z}{1+\beta}\right)^n e^{-\frac{1}{2}\frac{\alpha_0z}{1+\beta}-mi\phi(\omega-\omega_0)-ikz}
\end{equation}
The $m=0$ component of the series accounts for direct transmission. The next terms behave as the transmission factors of dispersive lines. Each dispersive line is characterized by the group delay:
\begin{equation} \tau_g^{(m)}(\omega)=m\mathrm{d}\phi(\omega-\omega_0)/\mathrm{d}\omega=m\mu(\omega-\omega_0)
\end{equation}
that grows linearly with $\omega$. In the following, for the sake of simplicity, $\tau_g^{(1)}(\omega)=\tau_g(\omega)$.

Let us concentrate on the first dispersive term, with $m=1$. The inverse group delay is nothing but the period of the absorption coefficient spectral modulation. Hence, the maximum size of the group delay is determined by the narrowest spectral feature one can engrave in the filtering medium.

Affected by different group delays, the successive dispersive terms can give rise to temporally separated contributions. This will be made clear in the following. If the successive contributions do not overlap in the time domain, the transmission efficiency of the $m$th term can be defined as:
\begin{equation}
\label{eq:eta_n}
\eta_m(\alpha_0z,\beta)=\left|\mathcal{T}_m(z,\omega)\right|^2.
\end{equation}
One easily verifies that:
\begin{equation}
\label{eq:total_transmitted_energy}
\sum_{m=0}^\infty\eta_m(\alpha_0z,\beta)= \mathrm{e}^{-\frac{\alpha_0z}{1+\beta}}\mathrm{I}_0\left(\frac{\beta\alpha_0z}{1+\beta}\right)
\end{equation}
where $I_0(x)$ stands for the $0$-order modified Bessel function. The $m=1$ response  efficiency reads as:
\begin{equation}
\label{eq:eta_1}
\eta_1(\alpha_0z,\beta)=\left(\frac{1}{2}\frac{\beta\alpha_0z}{1+\beta}\right)^2 \mathrm{e}^{-\frac{\alpha_0z}{1+\beta}}
\end{equation}
The maximum efficiency for the first-order response, 13.5\%, is reached at $\alpha_0z=4$, $\beta=1$. The same fraction of the incoming energy is carried away by the direct transmission component:
$\eta_0(\alpha_0z,\beta)=e^{-\frac{\alpha_0 z}{1+\beta}}$.
According to equation~(\ref{eq:total_transmitted_energy}), all the higher order terms collect less than 4\% of the incoming energy. Delayed transmission through a spectrally periodic sinusoidal filter is commonly known to be limited to 13.5\% efficiency. Not surprisingly, the locally periodic approximation leads here to the same figure.

Dispersive filtering near atomic resonance has already been used in combination with a time lens for real-time spectral analysis~\cite{menager01,crozatier04,crozatier06} and analog arbitrary waveform generation~\cite{renner07,damon10} in the prospect of RADAR applications.

When addressed in the context of atomic resonance, both the absorbing material programming and the signal filtering are usually identified as parts of a stimulated photon echo process~\cite{linget13}. Photon echo is known as a four wave mixing process, and it may be misleading to refer to non-linear optics when one really deals with linear filtering. The first two pulses of the photon echo sequence are just used to program the filter. Reference to photon echo might also restrict the filter programming to the canonical two pulse sequence. As a matter of fact, a vast variety of techniques has been used to program the material, as illustrated by various demonstrations ranging from the accumulated photon echo~\cite{hesselink1979,hesselink1981} to the rainbow analyzer~\cite{lorgere2002} and the atomic frequency comb (AFC)~\cite{nilsson2004,afzelius2009,afzelius2010}. Because of the flexibility of the programming step, we purposely let aside the specific preparation technique, concentrating on the most general features that can give us some insight on ultimate properties and limitations.

\subsection{Application to time reversal}
In this section we derive the image of an incoming waveform through our approximate imaging scheme in a single step, following the calculation achieved by J. Aza\~na \emph{et al.} in the spatial domain~\cite{azana2005shadow} and in the time domain~\cite{azana2005}.

Let the waveform $S(t)$, centered at $t=0$, be upconverted on the frequency chirped optical carrier $\mathcal{C}(t)$, resulting in an input field $\mathcal{E}(z=0,t)=S(t)\mathcal{C}(t)$.
The carrier field reads as:
\begin{equation}
\label{eq:input_carrier}
\mathcal{C}(t)=\mathcal{C}_0\mathrm{e}^{i\omega_1t-\frac{i}{2}rt^2}
\end{equation}
The instantaneous carrier frequency $\omega_1-rt$ equals $\omega_1$ at $t=0$ and decreases with time. Within the waveform duration $T$, the carrier frequency satisfies the dispersive filter operation condition $\phi(\omega-\omega_0)\gg1$, which leads to:
\begin{equation}
rT/2\ll\omega_1-\omega_0
\end{equation}
Propagating through the $m=1$, $L$-thick dispersive filter, the input field $\mathcal{E}(z=0,t)$ undergoes the transformation:
\begin{equation}
\tilde{\mathcal{E}}(L,\omega)=\kappa\mathrm{e}^{-\frac{i}{2}\mu(\omega-\omega_0)^2-ikL}\tilde{\mathcal{E}}(0,\omega)
\end{equation}
where $\kappa=\sqrt{\eta_1(\alpha_0L,\beta)}$, and where the time-to-frequency Fourier conjugation is defined by:
\begin{equation}
\tilde{f}(\omega)=\int\mathrm{d}tf(t)\mathrm{e}^{-i\omega t},\ f(t)=\frac{1}{2\pi}\int\mathrm{d}\omega\tilde{f}(\omega)\mathrm{e}^{i\omega t}.
\end{equation}

Let $r\mu=2$, which is the matching condition for time reversal with unit magnification. A tedious but straightforward calculation leads to:
\begin{equation}
\label{eq:transmitted_field}
\mathcal{E}(L,t)=\kappa\mathcal{C}_0\mathrm{e}^{i\psi(t)}\frac{1}{2\pi}\int\mathrm{d}\omega\tilde{S}(\omega)\mathrm{e}^{-i\omega\left[t-\tau_g(\omega_1)\right]+\frac{i}{2}\mu\omega^2}
\end{equation}
where:
\begin{equation}
\label{eq:carrier_phase}
\psi(t)=\omega_1 t+\frac{1}{2}r\left[t-\tau_g(\omega_1)\right]^2-\frac{1}{4}r\tau_g(\omega_1)^2
\end{equation}
If, in the integrand of equation~(\ref{eq:transmitted_field}), $\mu\omega^2\ll2\pi$ over the waveform bandwidth, the sum over $\omega$ reduces to the inverse Fourier transform of $\tilde{S}(\omega)$ at time $-t+\tau_g(\omega_1)$ and the transmitted field can be expressed as:
\begin{equation}
\label{eq:transmitted_field_final}
\mathcal{E}(L,t)=\kappa\mathcal{C}_0\mathrm{e}^{i\psi(t)}S\left(-t+\tau_g(\omega_1)\right)
\end{equation}
As pointed out above, the ignored phase contribution, $\mu\omega^2$, reflects the imaging imperfection of the approximate imaging scheme.


\begin{figure}[!htb]
\centering\includegraphics[width=12.5cm]{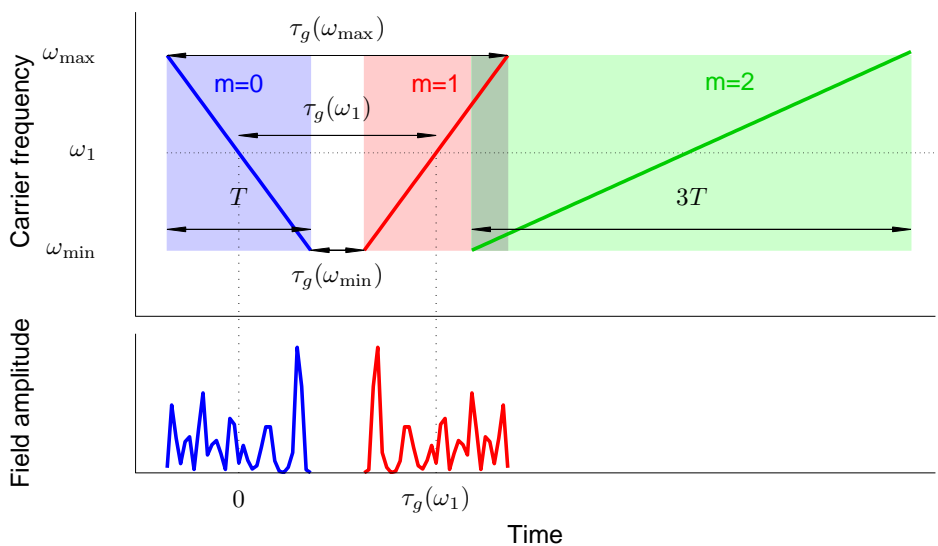}\caption{Carrier and signal time diagrams. The incoming carrier instantaneous frequency decreases with time from $\omega_{\mathrm{max}}$ to $\omega_{\mathrm{min}}$. On the contrary, the first-order response carrier frequency increases with time from $\omega_{\mathrm{min}}$ to $\omega_{\mathrm{max}}$. The $(\omega_{\mathrm{min}},\omega_{\mathrm{max}})$ interval is centered at $\omega_1$. The outgoing response, centered at $t=\tau_g(\omega_1)$, is time-reversed with respect to the incoming signal, centered at $t=0$. The carrier frequency range $[\omega_{\mathrm{min}},\omega_{\mathrm{max}}]$ and the signal duration $T$ are connected by
$2T = \tau_g(\omega_{\mathrm{max}})-\tau_g(\omega_{\mathrm{min}})$. In the figure, the first $(m=1)$ and second $(m= 2)$ order responses partly overlap. To avoid this, the signal duration must satisfy the condition $T\leq\tau_g(\omega_1)/2$.
}
\label{fig:time_reversal}
\end{figure}

The time-reversed waveform $S\left(-t+\tau_g(\omega_1)\right)$ is carried by the field $\mathcal{C}_0\mathrm{e}^{i\psi(t)}$. According to Eqs.~(\ref{eq:input_carrier}) and (\ref{eq:carrier_phase}), the sign of the carrier chirp rate at the output is reversed with respect to the input (see figure~\ref{fig:time_reversal}).

Finally the restored waveform should not be spoilt by higher order responses, as noticed in section~\ref{sec:dispersive_filter}.
As illustrated in figure~(\ref{fig:time_reversal}), the first order response ends at time $t_{max}^{(1)}= \tau_g(\omega_1) +T/2$. The second
order response starts at $t_{min}^{(2)}= \tau_g^{(2)}(\omega_1) -3T/2= 2\tau_g(\omega_1) -3T/2$. Those responses do not overlap
temporally, provided $ t_{max}^{(1)} \leq t_{min}^{(2)} $, which reduces to: $T\leq\tau_g(\omega_1)/2$. The carrier frequency range $[\omega_{\mathrm{min}},\omega_{\mathrm{max}}]$ and the signal duration are connected by $2T = \tau_g(\omega_{\mathrm{max}})-\tau_g(\omega_{\mathrm{min}})$.

\section{Experimental}\label{sec:experimental}
In this section we present the experimental implementation of time-reversal with the approximate scheme described in Sec~\ref{sec:approximate_imaging}, where the dispersive filter is created in an absorbing medium.

\subsection{Preparation of a dispersive filter}
\label{sec:exp_dispersive} Famous for its absorption line in the telecom wavelength range, the erbium Er$^{3+}$ ion inserted in yttrium orthosilicate (Y$_2$SiO$_5$) offers spectroscopic properties allowing preparation of a dispersive filter with a large dispersive power. On the one hand, each individual Er ion has a homogeneous linewidth of a few kHz when cooled down at $2$~K in a 2-tesla magnetic field. On the other hand, environment inhomogeneity at the ion sites induces an inhomogeneous distribution of the individual lines, resulting in a GHz-wide absorption profile for the ensemble. Shining a spectrally narrow laser on such a profile transfers resonant ions to their excited state and creates a long-lived dip in the absorption profile. This spectral hole-burning process is at the heart of the preparation of our dispersive filter.

\begin{figure}[!htb]
\centering\includegraphics[width=9cm]{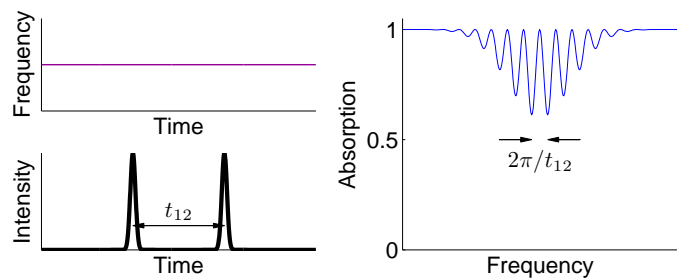}\caption{Two monochromatic pulses separated by $t_{12}$ induce a sinusoidal modulation of the absorption profile with spacing $2\pi/t_{12}$. }
\label{fig:grating1}%
\end{figure}

\begin{figure}[!htb]
\centering\includegraphics[width=9cm]{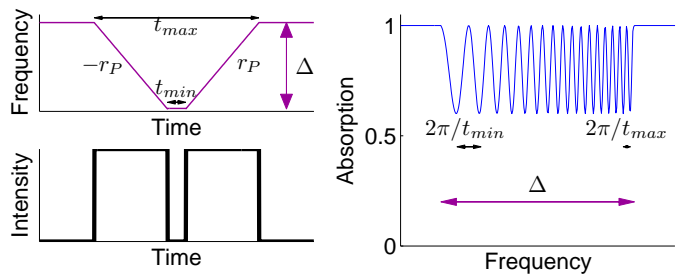}\caption{Two chirped pulses with opposite chirp rates $-r_P$ and $r_P$ induce a quasi-sinusoidal absorption grating with a variable spacing. }
\label{fig:grating2}%
\end{figure}

The preparation of the filter is inspired from the usual stimulated photon echo scheme. The first two pulses $\mathcal{E}_1(t)$ and $\mathcal{E}_2(t)$ of a stimulated photon echo sequence burn a sinusoidal spectral grating in the absorption profile, with a spacing equal to the inverse time separation of the two pulses  (figure~\ref{fig:grating1}). One can extend this observation to a pair of chirped pulses with opposite chirp rates $-r_P$ and $r_P$: each homogeneous frequency class of atoms senses a pair of pulses with a frequency-specific time separation~\cite{crozatier04,crozatier06}. This way, the inhomogeneously broadened profile is shaped into a quasi-sinusoid with linearly varying period, hence the dispersive filter with $\mu=2/r_P$ (figure~\ref{fig:grating2}).

The crystal used in our experiments is a $10$~mm-long $0.005\%$ Er:YSO crystal, cooled down to $2$~K in a helium bath cryostat. A $2$-tesla magnetic field is applied by means of superconducting split coils in the sample space.

\begin{figure}[!htb]
\centering\includegraphics[width=12cm]{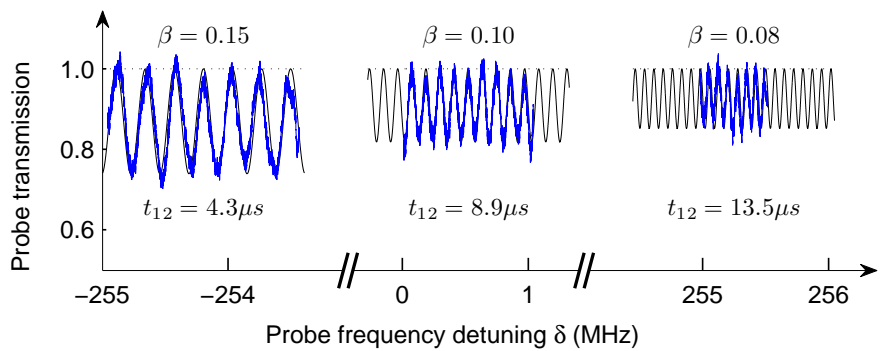}
\caption{Normalized transmission measurement around different frequencies within the $1.09~$GHz engraved frequency range (blue line). After the two programming pulses, a weak slowly chirped pulse is sent to the crystal and its transmitted intensity is monitored. Depending on the spectral range probed, one observes a locally sinusoidal modulation of the transmission, with frequency-dependent spectral period $1/t_{12}$ and contrast $\beta$. Thin black line: $[1+\beta\sin(2\pi\delta t_{12} +\theta)]/(1+\beta)$.}
\label{fig:exp_filter}%
\end{figure}

We prepare the dispersive filter with two oppositely chirped pulses sent to the Er:YSO crystal, sweeping their frequency over $\Delta/2\pi=1.09$~GHz in $6~\mu$s, resulting in a chirp rate $r_P=1.14\times 10^{15}$~rad.s$^{-2}$. We probe the modified absorption profile by shining an attenuated, slowly chirped laser pulse ($2~$MHz in $40~\mu$s) around three frequencies (see figure~\ref{fig:exp_filter} for details). We use a single laser source for the preparation pulses and the probe pulse, namely an agile extended cavity diode laser that has been designed to perform reproducible GHz-wide frequency sweeps within a few $\mu$s without mode hops~\cite{crozatier07}. The laser frequency sweeps are obtained by feeding a linear voltage ramp to an intra-cavity electro-optic crystal, with smooth slope changes to avoid excitation of piezo-electric resonances in the crystal. The laser is fed into an erbium-doped fiber amplifier and shaped in amplitude with a free-space acousto-optical modulator.

The transmitted probe intensity is shown in figure~\ref{fig:exp_filter}. As expected, we observe a locally sinusoidal modulation with frequency-dependent spacing, corresponding to the pulse separation sensed by the atoms absorbing at that frequency. The resulting dispersive power of such a filter is $\mu=2/r_P=1.5$~ms/nm, that is $8$ orders of magnitude larger than the dispersive power of a standard km-long fiber at $1.55~\mu$m ($17$~ps/nm/km). The measured contrasts $\beta$ range from $0.08$ to $0.15$.

We insist on the fact that the $13.5~\%$ maximum efficiency announced in section~\ref{sec:dispersive_filter} corresponds to a quasi-sinusoidal grating with unity contrast and a maximum optical depth $\alpha_0L=4$. According to equation~\ref{eq:eta_1}, the efficiency in the first order response varies like $\beta^2$. With an opacity $\alpha_0 L=2$ and based on the contrast measurements shown in figure~\ref{fig:exp_filter}, we expect a retrieval efficiency $\eta_1\simeq1.3$\textperthousand.

The low contrast achieved in our experiments originates from the perturbative character of the preparation step. However, thanks to its short duration ($\simeq15~\mu$s), the preparation step can be repeated many times within the grating lifetime, similarly to accumulated photon echoes~\cite{hesselink1979, hesselink1981}. This should increase the grating contrast and therefore enhance the efficiency, up to $13.5\%$. Besides, optimization of the grating shape is known to allow even higher efficiencies beyond the $13.5\%$ limit. For example, an efficiency close to $20\%$ has been demonstrated with a rectangular grating in Tm:YAG, in the context of quantum memories~\cite{bonarota2010}.

\subsection{Time-reversal of a $5~\mu$s-long signal}

\begin{figure}[!htb]
\centering\includegraphics[width=12cm]{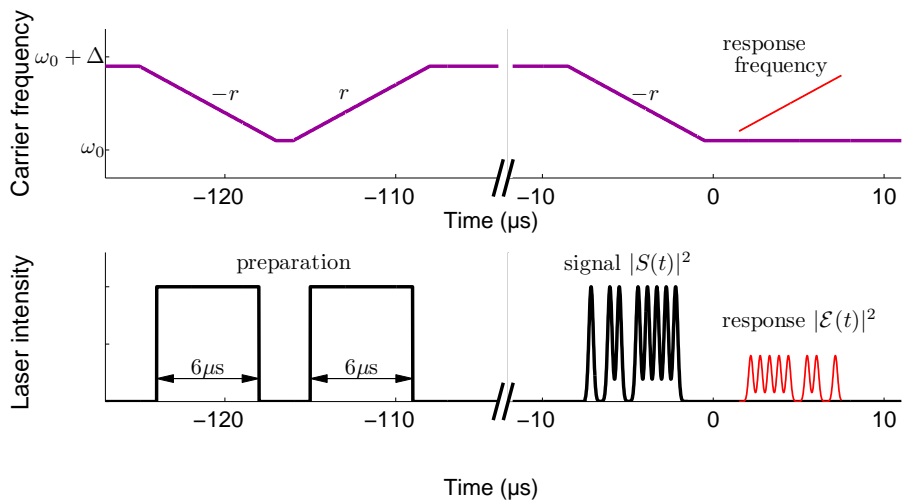}
\caption{Experimental pulse sequence. Thick lines: incoming preparation and signal pulses. Thin line: dispersive filter first order response, emitted as a chirp with rate $r$, carrying the time-reversed signal}
\label{fig:pulse_seq}%
\end{figure}

The complete pulse sequence is shown in figure~\ref{fig:pulse_seq}. To ensure time-reversal with unit magnification, the signal $S(t)$ is upconverted on a chirped optical carrier with chirp rate $r_S=-r_P=-r$ with an acousto-optic modulator. The signal duration ($5~\mu$s) is compatible with typical radar signal durations. The signal propagates through the dispersive filter programmed in the Er:YSO crystal by the preparation pulses. The first order response of the filter is emitted as a chirped sequence of pulses with rate $r$.

We use the  experimental setup described in section~\ref{sec:exp_dispersive} to generate the whole sequence. An additional acousto-optical modulator attenuates the strong pulses transmitted through the Er:YSO crystal, to protect the avalanche photodiode Thorlabs APD110C used for detection.

We emphasize that the absorption profile of our Er:YSO crystal has a full width at half maximum of $2.7~$GHz, and its optical depth $\alpha_0 L$ is not constant across the $1.09~$GHz dispersive filter width. The filter contrast $\beta$ is not constant either across the filter width, as shown in figure~\ref{fig:exp_filter}. A variation of the efficiency in the $m=0$ and $m=1$ responses should therefore be expected [see equations~(\ref{eq:T_n}) and (\ref{eq:eta_n})].

Given the measured  grating contrast ($\beta\simeq10~\%$) and typical optical depth in our sample $\alpha_0 L=2$, we expect a first-order response efficiency $\eta_1\simeq1.3$\textperthousand, and a second order response much less efficient than the first order one: $\eta_2=2\times 10^{-3} \eta_1$. This is why we allow a time gap between the signal and the echo that makes the second order partly overlap with the first order response (figure~\ref{fig:time_reversal}).

Due to causality, the whole input signal must be confined before the time-reversed response is emitted. During this time, the information is saved in the crystal as a superposition state that is sensitive to decoherence. The various mechanisms at play are described in~\ref{sec:decoherence}. With our experimental parameters, the decoherence mechanisms remain small enough and the time-reversal efficiency is not affected.

The time-bandwidth product $P$ of our time-reversal protocol is limited by the intrinsic properties of the atomic ensemble. Indeed, $P$ reads as: $T\times\delta$ where $T$ is the input signal duration and $\delta$ is its spectral width. The Fraunhofer condition imposes that $P\ll\sqrt{\Delta T/2}$. $\Delta$ cannot exceed the inhomogeneous linewidth of the atomic ensemble. With a signal duration of $5~\mu$s, this leads to $P \ll 80$. Using a different crystalline host such as LiNbO$_3$, where the Er inhomogeneous linewidth reaches $180$~GHz and the decoherence effects are similar to that in YSO~\cite{thiel-JL01}, would allow a larger bandwidth and a larger time-bandwidth product for our time-reversal protocol. This would require a laser able to perform fast reproducible sweeps over such a frequency range. However, as long as approximate imaging is used, the Fraunhofer condition remains.

\subsection{Exploring the Fraunhofer condition}

In this section we investigate the signal distortion due to the violation of the Fraunhofer condition. We explore the effect of the Fraunhofer condition by sending signals with different bandwidths on the dispersive filter described above. All the incoming signals consist in a train of Gaussian pulses $\sum_i \exp\left[-\frac{(t-t_i)^2}{2\tau^2}\right]$ with $\tau$ ranging from $18$ to $67$~ns. Successive peaks are separated by multiples of $207$~ns. The Fraunhofer condition is satisfied when $\tau\gg\sqrt{2}/\left(2\pi\sqrt{r}\right)=17$~ns.

\begin{figure}[!tb]
\centering\includegraphics[width=12cm]{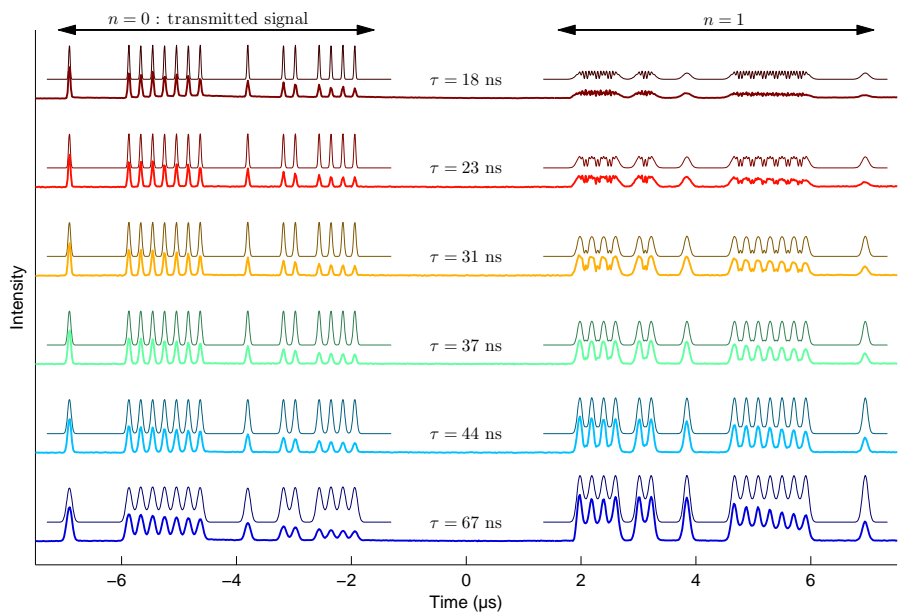}
\caption{Around the Fraunhofer limit. Thick lines: experimental transmitted signal and first-order response measured on the same detector after propagation through the dispersive filter described in figure~\ref{fig:exp_filter}. The input signal consists in a train of identical Gaussian pulses with typical widths $\tau$ ranging from $18$ to $67$ ns. The variable intensity of the experimental $m=0$ and $m=1$ responses is due to the variation of optical depth and filter contrast across the filter frequency range. Thin lines: input signal and calculated response intensity of an ideal dispersive filter with dispersive power $\mu=1.5$~ms/nm, only adjusted in height to be compared easily with the experimental signal. The curves are vertically offset for clarity.}
\label{fig:fraunhofer}%
\end{figure}

Figure~\ref{fig:fraunhofer} shows the transmitted signals and their first-order response detected on the same avalanche photodiode. The transmitted signal is the $n=0$ response of the dispersive filter. Its intensity varies because of the crystal optical depth $\alpha_0 L$ and filter contrast $\beta$ variation over the filter frequency range, as expected from our analysis of figure~\ref{fig:exp_filter}.

The $m=1$ response contains the time-reversed counterpart of the input signal. In these experiments, the  maximum efficiency of the time-reversal process reaches $1.6$\textperthousand, in agreement with the measured population grating contrast. Like the $m=0$ response, the first-order response intensity varies over the signal duration because the optical depth $\alpha_0 L$ and the filter contrast $\beta$ are not constant over the filter frequency range.

\begin{figure}[!tb]
\centering\includegraphics[width=10cm]{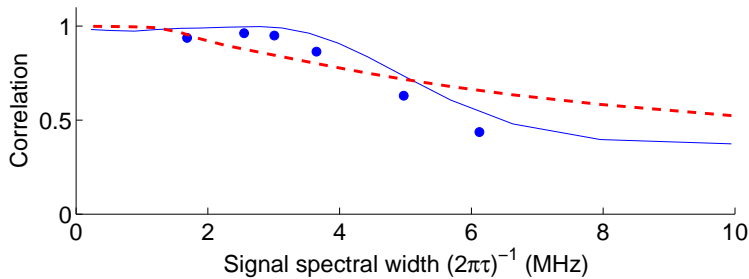}
\caption{Fidelity of the time-reversal with respect to the signal spectral width. Circles (resp. solid line): correlation coefficient between the input signal and its experimental (resp. calculated) time-reversed replica shown in figure~\ref{fig:fraunhofer}. Dashed line: correlation coefficient between the input signal and the calculated equivalent pinhole camera image.  The two setups offer the same bandwidth, with a cutoff around $(2\pi\tau)^{-1}=4~$MHz, ie $\tau=40~$ns.}
\label{fig:fraunhoferc}%
\end{figure}

The signal is faithfully reversed as long as $\tau\geq40$~ns, in agreement with the Fraunhofer condition $\tau\gg 17$~ns derived above. When the signal spectral width exceeds the filter bandwidth, each peak of the time-reversed signal is stretched and overlaps with its nearest neighbours that are emitted at different frequencies, hence the beatnote between close pulses. As expected, high frequency components in the input signal induce strong distortion in the filter response. We compare the experimental measurements with the calculated signals as given by equation~(\ref{eq:transmitted_field}) and see almost perfect matching. We visualize the fidelity of the time-reversal by plotting the correlation coefficient between the input signal and its experimental and calculated time-reversed replicas in figure~\ref{fig:fraunhoferc}.

\subsection{Simulation of a pinhole time camera}
\begin{figure}[!htb]
\centering\includegraphics[width=12cm]{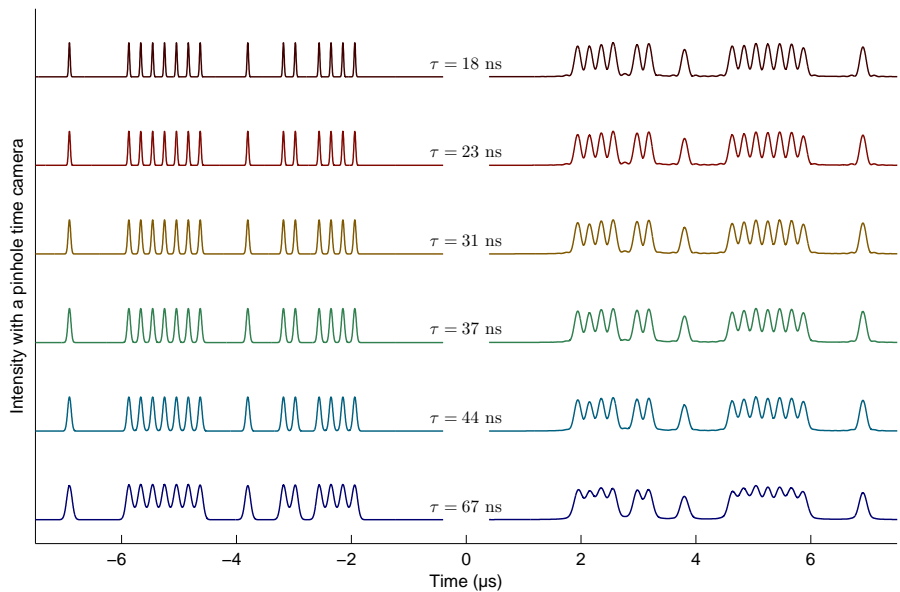}
\caption{Simulation of a pinhole time camera. Left curves: input signals (identical to those used in our time-reversal experiments). Right curves: calculated pinhole response. The curves are vertically offset for clarity.}
\label{fig:fraunhofer_pinhole}%
\end{figure}

We calculate the image obtained with a time pinhole camera setup~\cite{kolner97}, with two successive dispersive lines with $\mu=1/r=0.75~$ms/nm on either side of a rectangular time window. We set the pinhole time window to $\zeta=40$~ns in order to make the pinhole camera and the time-reversal setups comparable. Note that the Fraunhofer condition is satisfied inside the dark chamber. In the conventional optics domain, incoherent light is needed at the input of the pinhole camera to allow some light to reach the pinhole even when coming from the outer part of the object. In our time pinhole camera simulation, we introduce incoherence in the input signal by multiplying it with a random phase distribution factor. The calculated outputs are averaged over $500$ random phase distributions, normalized and shown in figure~\ref{fig:fraunhofer_pinhole}. One recognizes the pinhole outputs as the convolution of the time-reversed input signals with the impulse response of the pinhole setup, ie the Fraunhofer diffracted image of the pinhole aperture: $\rm sinc^2 \left( r\zeta t/2 \right)$. This impulse response has a full width at half maximum of $125~$ns, larger than or comparable to the input signal details, which results in a broadening of the signal details and a contrast reduction in all the time-reversed images shown in the figure.

In figure~\ref{fig:fraunhoferc} we display the fidelity of time-reversal with a pinhole time camera. Perfect time-reversal with a pinhole occurs only when the input signal details are larger than the impulse response typical temporal width, ie when the signal spectral width $(2\pi\tau)^{-1}\ll 1.3~$MHz. Beyond this value, the pinhole camera fidelity decreases at a slower pace than our approximate time-reversal scheme. Indeed, beyond the Fraunhofer limit, high-frequency components are simply cut-off by the pinhole, unlike in the time-reversal experiment where they strongly distort the reversed signal. The two architectures nevertheless share a similar bandwidth, close to $\sqrt{2/r}$.

\section{Conclusion}
To sum up, we have demonstrated the time-reversal of optically carried signals with an approximate temporal imaging scheme, involving a time lens and a dispersive filter. The dispersive filter is obtained with a rare-earth-ion-doped crystal whose absorption profile has been modified with spectral hole burning. The dispersive power of our filter exceeds that of conventional fibers by many orders of magnitude. Because of its approximate character, our time-reversal scheme suffers from a bandwidth limitation, analog to the Fraunhofer condition in optics. We investigate this condition experimentally, and observe strong signal distortion when the condition is not satisfied. We connect our time-reversal scheme with an equivalent  time domain pinhole camera. We observe that although they share the same bandwidth, they distort the input signals differently.

For better time reversal bandwidth, taking full advantage of the filter bandwidth, one cannot dispense with true temporal imaging, which involves a time lens and two dispersive lines, respectively located upstream and downstream from the lens. Those elements are needed to conjugate the temporal object and its time-reversed image. To operate with two dispersive elements, the time lens must deal with an optical signal at the input and the output. A wave-mixing-based non-linear optical device commonly satisfies such conditions~\cite{bennett94,bennett99,bennett00,foster09}. When the dispersive filters are programmed in an absorbing medium, one is faced with the additional requirement of having both filters work at the same wavelength. Then the time lens shall be based on difference frequency generation (DFG) rather than on more usual sum frequency generation (SFG). For instance, entering the lens at 1.5 $\mu$m, the signal shall emerge at the same wavelength if the lens works as a DFG-mixer with a chirped field at 750 nm. Those non-linear techniques are well documented~\cite{chou1999}.

More problematic is passing the signal twice through the shaped-absorption-based filter. According to  section~\ref{filter_in_abs_medium}, single pass optimized efficiency does not exceed 13.5\%, which leads to less than 2\% efficiency through two successive dispersive filters. However, we know that the 13.5\% barrier can be broken. For instance, optimizing a periodic filter in Tm$^{3+}$:YAG, we have been able to demonstrate about 20\% efficiency with rectangular-shaped spectral grooves in the context of quantum memory for light~\cite{bonarota2010}.

\ack
The authors thank V. Crozatier and V. Damon for their contributions at earlier stages of the experiment, and L. Morvan for his expertise relative to radar signal processing.
The research leading to these results has received funding from the People Programme (Marie Curie Actions) of the European Union's Seventh Framework Programme FP7/2007-2013/ under REA grant agreement no. 287252.

\appendix

\section{Sources of decoherence}
\label{sec:decoherence}
The efficiency of our time-reversal filter is affected by decoherence mechanisms, originating from the finite homogeneous linewidth, but also from spectral diffusion and instantaneous spectral diffusion~\cite{Bottger-PRB06}.

Spectral diffusion refers to the fluctuations of each ion transition frequency as a result of interaction with the environment. In crystals doped with Kramers ions, such as Er$^{3+}$, spectral diffusion is mainly caused by magnetic dipole-dipole interaction between the active ions themselves. This contribution can be strongly reduced by an external intense magnetic field that locks the spins in a fixed orientation. In Er:YSO we apply a 2-tesla field in the $(D_1,D_2)$-plane, $135^\circ$ from the $D_1$-axis~\cite{bottger09}, while maintaining the sample at 2K. The subsisting spectral diffusion affects three pulse photon echoes (3PE) in two different ways.

On the one hand, spectral diffusion destroys the atomic coherence that carries optical excitation during the $t_{12}$ interval between the first two pulses. Indeed the frequency fluctuations alter the relative phase of the atomic states in a cumulative way. The same coherence deterioration takes place between the third pulse and the echo.

On the other hand, spectral diffusion tends to erase the periodic spectral structure that is engraved in the absorption profile by the first two pulses. The signal is the more affected as the structure period, given by $1/ t_{12}$, is narrower, and as the waiting time $T_W$ from the second to the third pulse is longer. However, the decay saturates as a function of $T_W$ since the frequency excursion range is limited.

In Er:YSO, the observed 3PE intensity $t_{12}$-dependent decay is consistent with $\exp[-4\pi t_{12} \Gamma_{eff}(t_{12},T_W)]$ where the effective decay rate $\Gamma_{eff}(t_{12},T_W)$ reads as~\cite{Bottger-PRB06}:
\begin{equation}
\Gamma_{eff}(t_{12},T_W)= \Gamma_0 + \frac 1 2 \Gamma_{SD} [R t_{12}+(1-\exp( -R T_W))].
\end{equation}
In this expression $\Gamma_0$, $\Gamma_{SD}$ and $R$ respectively represent the single ion homogeneous width, the frequency fluctuation range and the inverse correlation time of the fluctuations. The $T_W$-dependence saturates when $R T_W>1$. In our crystal we have measured $\Gamma_0 = 800$~Hz, $\Gamma_{SD} = 3~$kHz, and  $R=10\pm1\times10^3$~s$^{-1}$.

Our time-reversed signal can be analyzed as a combination of 3PE sequences spreading over a range of different $t_{12}$ and $T_W$ values. With a minimum $T_W$ value of $\sim100~\mu$s we operate in the saturation regime, and with $t_{12} <6~\mu$s we get $\exp[-4\pi t_{12} \Gamma_{eff}(t_{12},T_W)] > 0.93$.

Instantaneous spectral diffusion refers to random frequency shifts of each ion caused by excitation-induced changes in the ions environment. As ions in the lattice are excited by the incident light, changes in the excited ions properties alter the resonance frequencies of nearby ions. Based on~\cite{liu90} we estimate that the preparation pulses induce a broadening of the single ion homogeneous linewidth $\Gamma_{0}$ up to $80$~kHz, when all the erbium ions of the filter frequency range are transferred to their excited state. In our experiments, we minimize the contribution of the instantaneous spectral diffusion by choosing a large beam waist diameter ($130~\mu$m) in the crystal, a moderate beam power ($5$~mW), and a rather short preparation duration (two $6~\mu$s-long pulses).


\bibliographystyle{iopart-num}
\section*{References}

\providecommand{\newblock}{}

\end{document}